# A real-time QKD system based on FPGA

Hong-fei Zhang, Jian Wang, *IEEE member*, Ke Cui, Chun-li Luo, Sheng-zhao Lin, Lei Zhou,
Hao Liang, Teng-yun Chen, Kai Chen, Jian-wei Pan

*Abstract*—A real-time Quantum Key Distribution System is developed in this paper. In the system, based on the feature of Field Programmable Gate Array (FPGA), secure key extraction control and algorithm have been optimally designed to perform sifting, error correction and privacy amplification altogether in real-time. In the QKD experiment information synchronization mechanism and high-speed classic data channel are designed to ensure the steady operation of the system. Decoy state and synchronous laser light source are used in the system, while the length of optical fiber between Alice and Bob is 20 km. With photons repetition frequency of 20 MHz, the final key rate could reach 17 kbps. Smooth and robust operation is verified with 6-hour continuous test and associated with encrypted voice communication test.

## I. INTRODUCTION

Quantum Key Distribution (QKD) is an ideal way for solving the key distribution problem. The system can creates an unconditionally secure key string between Alice and Bob (two sides of QKD system) by using quantum states as a carrier for distributing secure keys. With the development of the quantum cryptography in the latest 20 years, implementations of QKD system have got more and more attentions. Therefore, various protocols have been put forward, such as BB84 protocol [1], B92 protocol [2], EPR protocol [3] and so on. Meanwhile, there arises a number of implementation schemes, and even for network set-up schemes.

The security of BB84 protocol was critically validated by Shor and Preskill in 2000 [4]. And afterwards, BB84 protocol is improved to solve some technical difficulties in a practical application, serving as decoy state protocol [5-7]. The decoy state protocol can prevent the system from photon number splitting (PNS) attack stemming from multiple photons in light source, expected to be an ideal single photon source. In recent years, quantum telephone network [8] and all-pass star network [9] have been accomplished through our participations. The QKD system introduced in this paper is based on real-time and

Manuscript received May 30, 2012; revised Aug. 3, 2012. This work was supported by Chinese of Academy of Science, the National Fundamental Research Program of China under Grant 2006CB921900, the National High Technology Research and Development Program (863 Program) of China under Grant 2009AA01A349, the Fundamental Research Funds for the Central Universities, National Natural Science Funds of China under Grant No: 11178020, 11175170 and the CAS Special Grant for Postgraduate Research, Innovation and Practice

The authors are with the Hefei National Laboratory for Physical Sciences at Microscale and Department of Modern Physics, University of Science and Technology of China, Hefei, Anhui 230026, China (Jian Wang is the corresponding author, e-mail: wangjian@ustc.edu.cn)

efficient secure key extraction algorithm through Field Programmable Gate Array (FPGA), and our system employs decoy state protocol and technique of wavelength division multiplexing (WDM) for the transportation of signal light and synchronous light in a single optical fiber. The focus points in this paper are real-time system control and secure key extraction of QKD system.

The QKD system is typically divided into 3 layers [10-11]: the physical layer at the base is to complete the communication of quantum channel and physical settings; the key-extraction layer in the middle is to ensure the algorithm of extraction; the key-application layer at the top contains methods of secure key applications such as secured telephone and video communication. In view of overall function, the QKD system can be divided into several parts including key extraction control, key extraction algorithm, key application, optics module, and single photon detectors (SPD). An all-in-one machine chassis is adopted in the QKD system stated herein, into which each communication party is integrated except the SPD. The layered structure of a typical QKD system is shown as Figure 1.

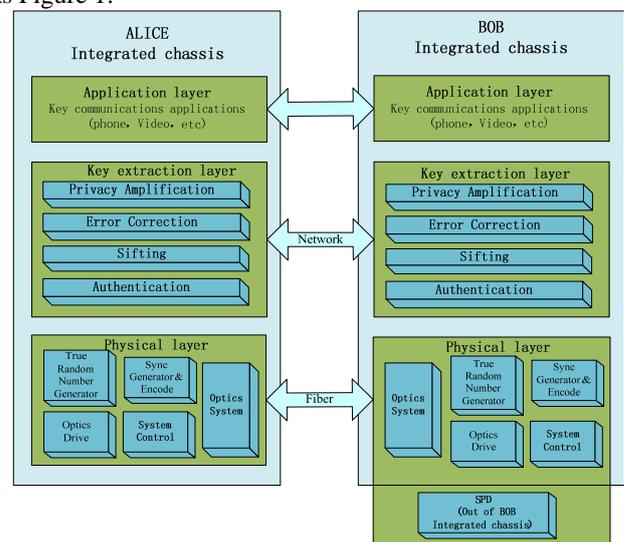

Figure 1 Diagram of layered structure of the QKD systems

Based on the clock jitter generated in FPGA, a True Random Number Generator (TRNG) with high-speed and low resource cost is designed in the physical layer to ensure the randomness of quantum signal, while *data bit/frame synchronization and encoding mo*dule to ensure the information synchronization in two-party communication. Besides, an overall *control* module is necessarily developed for the stable operation of QKD system, while the channel of USB and network are used as a classical channel for public information transmission.

Three major steps are taken in the middle key-extraction layer: (1) Sifting, Bob measures the basis information of the photon transmitted from Alice, and rules out non-identical quantum bits (qubits) from both parties so as to generate shifted keys containing some error bits due to the channel noise. (2) Error correction [12], Alice and Bob correct errors in sifted keys by exchanging some public information via the classic channel to achieve the uniformity of the keys for both. (3) Privacy amplification [13], some hash functions are used to shorten the corrected keys and reduce the amount of information obtained by Eve as much as possible.

Meanwhile, authentication is normally needed in a secure implementation of QKD. It hasn't yet been implemented in our current system, and we will add this function in the future work.

In this paper, the structural design of QKD system based on FPGA will be discussed in section two, its operational status and data analysis will be given in the section three and a conclusion will be presented in section four.

## II. SYSTEM DESIGN

### 2.1. System Overview

The QKD system herein uses the design of polarization encoding and the decoy state protocol, as well as a synchronization scheme of sync-light involving Alice and Bob. Between them there exists a quantum channel using optical fiber and a classic channel using network, and all of their major parts are integrated into an all-in-one chassis of quantum cryptography, containing transmitting function and receiving function. Therefore, the chassis of Alice and Bob are almost identical except that a set of Single Photon Detectors (SPDs) is needed on the Bob side.

An integrated quantum cryptography chassis can be divided into optical system, QKD control system and key extraction system. The structure of the optical system is shown as Figure 2. Four kinds of polarized light emitted from laser sources are coupled into one optical fiber through PBS and BS. The voltage-controlled polarization controller (PC) is for adjusting the polarization and controlling the system QBER based on the coordination of a QKD control system and a *high-voltage* module, which will be more discussed in 2.3.4.

The adoption of WDM in QKD system can facilitate the practical application by coupling the sync-light and signal light into one optical fiber. The wavelengths of sync-light and signal light are 1570 nm and 1550 nm, respectively. The SPD we used is based on InGaAs-type APD. The detection efficiency is about 12% and it's dark count rate is about $5*10^{-6}$/Gate.

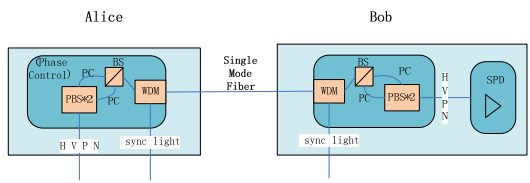

Figure 2 structure of the optical system

QKD control system and key extraction system are designed in the form of flashboard module and based on the core – an FPGA-based and easily-extended system control board. The board completes the function in physical layer and key extraction layer with the help of a *high-voltage* board, a *laser source* board, an optical module, a single-board-computer and a mother board for interacting between with these flashboards [14]. The function diagram is shown as the figure 3.

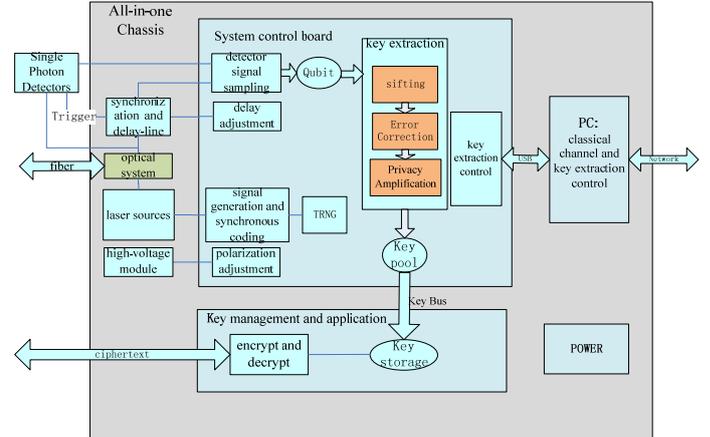

Figure 3 function diagram of the quantum cryptography integrated chassis

The system control board contains multiple function modules, such as *key extraction algorithm* module, *true random number generator* module, *signal generation and synchronous coding* module, *detector signal sampling* module, *delay adjustment and polarization adjustment* module; the single-board- computer and the control board are together cooperatively for the *key extraction control* module and *classical data channel* module. Besides, within the chassis there are the *laser source* module, *synchronization and delay-line* module, *high-voltage* module, *key management and application* module.

### 2.2 Physical Layer Control

In the physical layer there are a series of control processes from light emitting, signal synchronization to data encoding.

Firstly, the system must generate continuous random sequence to form quantum bits. The proportion of signal state: decoy state: vacuum state is 6:1:1 and H:V: P: N (four polarizations) is 1:1:1:1. The *signal generation and synchronous coding* module then encode the random sequence and store it according to the sequence, and meanwhile output the signal to drive lasers. The *laser source* module will generate the sync-light and the signal light according to the driving signal. The intensity of the signal state light must be adjusted to be three times of the intensity of the decoy state light in each polarization state. The average photon number of signal state pulse and decoy state pulse are 0.6 and 0.2, while the intensity of sync-light is much higher to make sure the information synchronization between Alice and Bob. Here the choice of the average photon number of 0.6 is roughly optimal according to the systematic theory of decoy state based quantum key distribution [5,6]. Moreover, our system is unconditional secure and is particularly robust against to the beam splitting attack, which inherits to the intrinsic advantage of decoy state schemes.

Secondly, the *synchronous and delay-line* module gets the sync-light signal and transmits it to the control board at the Bob's side, meanwhile delays the sync-signal and transmits it to the SPD as the gate-controlled signal. The *high-voltage* module and the *control* module are used to control polarization controllers to complete the polarization adjustment and narrow down system error rate to a proper range. The system control board gets the photon signals gathered by the *detector signal sampling* module at the Bob's side and encodes the data according to the received sync-signal.

Then the coding data in the two communication sides completes the secure key extraction algorithm through the data exchange in classic channel. At last the final keys are extracted and outputted to the *key management and application* module at both sides.

In the physical layer, high-speed TRNG and information synchronization mechanism are designed to ensure the randomness of polarization of emitted photons and the steady process of key extraction.

### 2.2.1 High-speed True Random Number Generator

Prior to the sifting, qubits extracted from the coding data need to be gained from two communication parties. Therefore a quality and efficient true random number generator and information synchronization mechanism are necessary preconditions to gain the high-quality qubits.

We have developed a high-speed true random number generator (TRNG) in FPGA [15]. The *TRNG* module generates random numbers based on jitter of high frequency clock, and the randomness source is fundamentally the thermal noise of the electronic system. We get ideal random numbers by using multiple sampling between the high frequency clocks and post-processing for the correction of bias. The random number generation speed of single *TRNG* module can reach 20 Mbps while the resource cost of FPGA is rather low, only 71 Logic Element. Five TRN*G* modules are used in one QKD system and the randomness of TRNG has passed the NIST and DIEHARD tests, thus guaranteeing true randomness of quantum signal and the security of system.

### 2.2.2 Information Synchronization Mechanism

The information synchronization between Alice and Bob is a crucial ingredient in the QKD system, and the completion of correct information synchronization ensures the subsequent proceeding of key extraction. In order to ensure the accuracy of photon location, we need to accurately transmit the sync-signal from Alice to Bob, and keep it synchronized with the output signals of the SPD. Better synchronization of signal can increase the accurateness of photon location, lower error rate and accordingly improve performance of the QKD system. If the synchronous is in the condition of small probability loss, the system can be resynchronized, which enhances the system's fault tolerance [16].

The implementation process of the information synchronization is shown as figure 4. The *sync-signal generation* module outputs frame format of sync-signal, and uses it as location information to encode the signal data. At the same time sync-signal is outputted to drive the sync optical laser. The emitted signal light and sync-light are coupled to one fiber, and are transmitted from Alice to Bob. The

*synchronization and delay-line* module at the Bob side receives, screens the sync-light, gets the sync-signal, before outputting it to the FPGA in the control board, meanwhile delays the sync-signal and outputs it to the SPD as the gate signal. The FPGA at Bob side receives the sync-signal, recovers the frame format of Alice, uses it as location information same as Alice to code the output data of the SPD. Then the coding data at the two communication sides are synchronous.

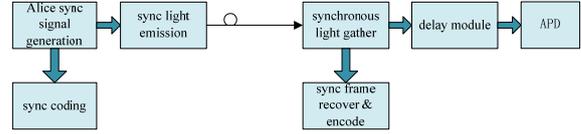

Figure 4 system information synchronization mechanism

The design of the frame format effectively solves the condition of small probability loss in sync-signal transmission process. The system clock for sync-signal generation is 20 MHz, the same as the quantum photon emission frequency. The sync-signal is generated by defining one frame as K state signal clocks and N low-voltage clocks, as shown in figure 5. Then the cycle length of each frame is a fixed value of K+N clocks, and thus the two communication sides will not get different frame count by judging the N low-voltage clocks. In our system, K is set to 1018 and N is 6, so the key generate rate will have very little loss of 6/1024. In each frame, the signal pulse is coded with 16 bits and 32 bits of frame head are used. If several clocks in one frame are lost in the transmit process, the coding data will be mismatched in this frame which could however be resynchronized at the next frame.

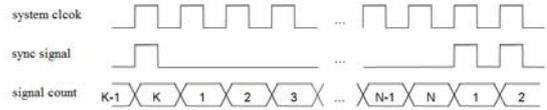

Figure 5 frame format of the sync-signal

The probability of losing sync-signal is very small in practical QKD system, bringing low wasting of large number of data due to frequent re-synchronization. However, sync-light and signal light received by Bob will have relative offset compared with Alice's because of offset caused by optical system and the speed difference between sync-light and signal light. As shown in figure 6, when Alice and Bob are correctly synchronized, the quantum bit error rate (QBER) will be much smaller than 50%, nonetheless if they are not properly synchronized, the QBER will approach 50%. Therefore, according to QBER, the adjustment of sync pulse count of Bob will obtain the offset from Bob to Alice and make them synchronized.

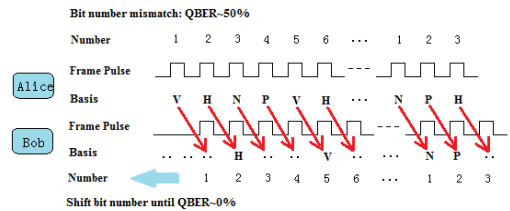

Figure 6 regulating of the sync-signal offset

### 2.3 Real-Time Key Extraction

Key extracting is an important step in QKD, no matter it's based on optical fiber or free space. Rapid and real-time key extraction is requisite in high-speed QKD system. In a practical system, efficient and real-time processing is needed to complete the sifting, error correcting, privacy amplification, which ensure the stable operation of QKD process.

The series of key extraction algorithms are mainly implemented by the control board based on FPGA while the single-board-computer provides many auxiliary functions like the system operation control and classical data exchange. The structure of system control board is shown as figure 7. Two cyclone III series FPGAs (EP3C120), which have built-in 4M Bits RAM and multi-16MBits SRAM as external storage resources are used for the completion of key extraction algorithm. A USB 2.0 high-speed serial data channel and network serve as classical data transmission channel to complete the data exchange between Alice and Bob.

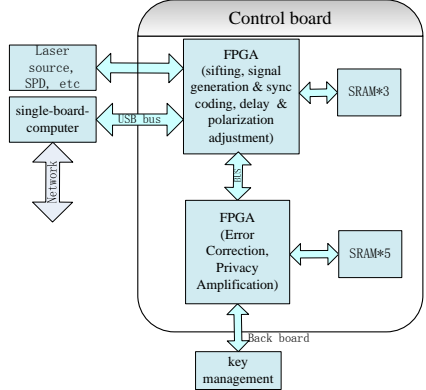

Figure 7 structure of the system control board

### 2.3.1 Sifting based on BB84 protocol

Sifting is a vital step in the process of key extraction, through which sifted key is generated with minor error bits between two communication parties. Only by ensuring fast stable and continued sifting process can the follow-up module receive sifted key for the error correction, privacy amplification algorithm ceaselessly and steadily generate the final key used for the secure application for key communication.

When Alice and Bob obtain the corresponding coding data, the *sifting* module can start the sifting operation according to the state information and basis information in the coding data. The sifting process can be summarized as the following three steps:

1.  Bob codes and stores the SPD output signal according to the sync-signal, and at the same time outputs the basis information of the coding data to Alice through the classical communication channel.
2.  Alice receives the coding data from Bob and decodes the data by frame format, and the synchronous encoding location information in the coding data from Bob will be obtained by Alice. Compare the basis bit of the coding data of the two sides, and retain the corresponding bit as sifted key of Alice.
3.  Bob receives the remaining data from Alice, finds the corresponding coding data stored in Bob's memory

according to the location information of the received data. Retain the corresponding bit as sifted key of Bob and calculate on the sampling error rate of decoy state and error rate of signal state.

The sifting algorithm module is processed in the FPGA of the system control board, whose logic structure is shown as figure 8. At first, the coding data generated by the *synchronous coding* module will be quickly deposited into SRAM on the control board, and the sifting algorithm can rapidly read the needed data according to the location information. The *USB module* exchanges data information with the other communication side through a USB chip to drive the running of the entire algorithm. The USB chip works as slave FIFO mode and four independent communication ports are used in the classical communication process. Two of them are uploading port for uploading the exchange data in the sifting process and the error correcting process. One of the ports downloads the exchange data in the sifting process and the other one downloads the exchange data in the error correcting process and the command data from the computer.

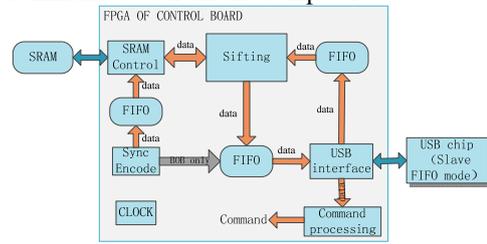

Figure 8 logic structure of the sifting algorithm

Compared with implementation through software, the real-time sifting algorithm completed in the FPGA has clear advantages with features of low data transfer, high-speed, stability and security.

Firstly, the photon repetition frequency of QKD system is 20 MHz. The coding data generation rate of Alice is 20M/s*4bits=80Mb/s with 2 bits indicating the four polarization states and the other 2 bits indicating the photon of being signal, decoy or vacuum state. The SPDs output number is about 1% of the 20 MHz emitted photons (include the effects of the photon number per pulse, the transmission loss and the SPD detection efficiency), so the coding data rate of Bob (R) is

$$R = F \cdot P \cdot A,$$
$$A = 16 + 32/(N \cdot P),$$

in which P=1%, F=20 MHz, A is the average bits used for encoding one detected data, and N is the numbers of pulse in one frame. So the coding data rate of Bob is about 20M/s*1%*(16+32/(1018*1%)) bits = 3.84 Mbps.

As the sifting is completed in FPGA, the FPGA will directly access the coding data in SRAM, and the *sifting* module doesn't need to upload the coding data of 80Mb/s generated at Alice side to the computer. The data need to be uploaded through USB is very few, only including coding data of Bob and the sifting retained data of Alice. This will save a lot of data transfer time between the FPGA and computer.

The USB chip has been tested whose ongoing communication bandwidth is around 240 Mbps. As the coding data from Alice doesn't need to be uploaded in real-time algorithm, the data traffic between the communication sides will be small. Thus the data exchange delay will be low, and the amount of coding data needed to be cached in Alice is small as well. Only two 16Mbits SRAMs are needed for the storage of coding data. However, if the sifting is performed in the software, the data transmission and the delay from its storage will demand numerous store space, as well as massive CPU occupancy for data reading and writing, which will definitely accompany with low efficiency of sifting algorithm implemented in the software.

The SRAM data access rate is 20M/s*16bis=320Mb/s in our system, so FPGA can access coding data of any address in 50ns and the sifting algorithm will complete the processing of one SPD output data in 20 clocks (1000ns). Therefore the data generated in 1s can be operated in 200ms by the algorithm, whose timeliness is unreachable via software-sifting.

Secondly, there is a minor chance of loss in high-speed USB data transmission, while no loss in direct data exchange between FPGA and SRAM, so the real-time data accessing in SRAM ensures the stability of QKD. Meanwhile it sets up physical isolation between system software and hardware, which avoids the risk of eavesdropping in data transfer process and strengthens the security in the key extraction process.

The *sifting* module is closely related to other system modules such as *sync coding* module, *laser sources* module, *synchronization and delay-line* module. Thus the *sifting* module needs to be operated under the supervision of the system *control* module. When the sifted key is extracted by the sifting module, the system *control* module will collect the count information of the SPD and the statistics information of the system error rate, and complete the state monitoring of the *sifting* module. When the change of the SPD count rate is rather obvious, the sifting process need to be paused for delay adjustment of trigger signal of the SPD. Moreover, the system error rate will drift with changes in environmental factors like time and temperature. When the error rate exceeds the preset threshold, the *polarization adjustment* module will start to work and decrease the error rate. The overall control of the system *control* module is a necessity to ensure the continuousness of the key extraction processing. More details on adjustment process will be discussed in section 2.3.4.

### 2.3.2 Real-Time Error Correction Algorithm

Errors in the sifted keys are inevitable because of the imperfect practical communication equipment, such as disturbance in the quantum channel and dark counts of single-photon detector. These errors must be eliminated by exchanging some information through public classic channel.

The Winnow protocol [17] introduces the idea of correcting errors using Hamming code, a linear block code, to decrease interaction times. Building on this idea, we develop a reconciliation protocol well-suited to be implemented on an FPGA [18].

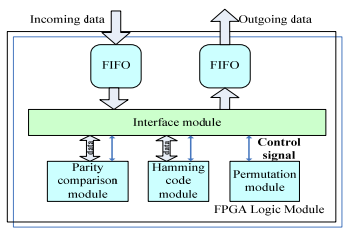

Figure 9 logic structure of the error correction algorithm

The reconciliation in our system adopts Hamming code as main correction tool that contains several iterations. In each of the iteration, the whole key string is divided into many segments according to current length (the length is a given value which is related to the error rate of the whole key string). Then Alice and Bob compute the parity of each segment and compare them. If the parity of a corresponding segment pair disagrees, then Hamming code is used to correct the error bit in this segment. Between every two neighboring iteration, one permutation is performed. When the reconciliation reaches the maximum number or it turns out all the parity of the segment pairs of both sides are same, CRC check is operated. If the CRC check passes then the reconciliation is realized successfully. The logic structure of the error correction algorithm is shown in figure 9.

In our design, the operating clock is set to 80 MHz. The efficiency of the algorithm is 1.4-1.7, and the sifted key can be processed with speed of 185 kbps. The total resource cost of FPGA is: logic element - 5734/119088(5%) and RAM - 393216/3981312(10%), which proves that resource cost is low. Taking advantage of the parallel feature of FPGA, one can easily integrate several *reconciliation* modules to improve the whole performance without doing more considerations.

### 2.3.3 Privacy Amplification

The corrected error rate of the key will decrease to almost zero, but the corrected key must be privacy amplified to make the Eve get less information. Through the privacy amplification, Eve can be detected by the unusual proportion of signal state and decoy state, meanwhile the *secure factor (SFactor)* will be reduced and the final key generation rate will decrease abnormally. We design real-time privacy amplification algorithm based on the FPGA, which has faster operation rate and better time feature compared with the conventional algorithm executed by the software on PC.

Our privacy amplification algorithm deals with the information after error correction by using the Toeplitz matrix [19]. The logic structure is shown as figure 10, and the implementation process includes the following steps:

1. The corrected keys are stored in RAM1 after the error correcting operation, and a set of pseudo random numbers are generated at the same time. When the amount of the gathered corrected keys increases to a processing unit of n bits (256 Kbits for example), privacy amplification process will start for once.

2. The *SFactor calculation* module on the single-board-computer calculates the SFactor for the corresponding corrected keys and transmits it to the *privacy amplification algorithm* module through the USB command port. Then the module will calculates the length of the final key (m bits) generated from the n bits corrected key according to the SFactor, and constructs a m rows and n columns Toeplitz matrix which have the same diagonal element by the generated random numbers. Our system stores the elements of the Toeplitz matrix into a on-chip memory (RAM2).

3. Take the n bits corrected key as a matrix of n rows and one

column, multiply by the Toeplitz matrix generated in step 2 and get a new matrix of m rows and one column. Then the m bits information will be the final keys after privacy amplification.

In the process of matrix multiplication, we use the concept of "block operation" because the matrix is very large, which divides the big matrix into many smaller blocks. The block can be parallel-processed for many bits at one time. For example, a matrix block is designed as a small 2 rows and 2 columns matrix, as shown in figure 11. The Toeplitz matrix and corrected key matrix can be divided into blocks respectively. The original Toeplitz matrix can be reconstituted to an M-rows and N-columns matrix by taking a block as an element of new matrix, in which M is equal to m/2 and N is equal to n/2. Multiplication of two blocks can be executed in one clock cycle due to the use of parallel arithmetic, so the entire matrix multiplication can be done in M · N clock cycles.

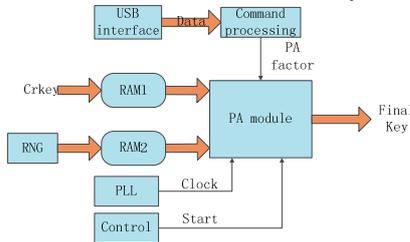

Figure 10 logic structure of the privacy amplification algorithm

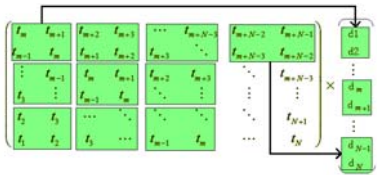

Figure 11 block operation of matrix

In the actual test of QKD, the block is set to a matrix of 40*40 bits and the working clock is 40 MHz. It needs four clocks to complete the reading and multiplication operation of one matrix block and thus it approximately needs 50M clocks (1.25s) to complete the privacy amplification process when the unit of privacy amplification algorithm is 256 Kbits and the SFactor is 0.3. The corrected key generation rate is about 70 kbps in the system, so the time is sufficient for the *privacy amplification algorithm* module to handle the corrected key.

Besides, the total resource cost of FPGA is low: logic element -1902/119088 (2%) and RAM - 656720/3981312 (16.5%). Thereby it's easy to integrate several modules to improve the data parallel processing speed. Compared with privacy amplification algorithm implemented in software, the algorithm based on FPGA displays the characteristic of parallel and greatly improves the speed of data processing.

### 2.3.4 Key Extraction Control

The key extraction control is primarily based on the single-board-computer in the integrated chassis and completed in the aid of the control board. Data exchange, control command and state-reporting data are transmitted through high-speed USB between single-board-computer and the control board [20].

Control software for key extraction is designed in the single-board-computer, which realizes the control of system command, QBER monitoring and continuing QKD process, data exchange and state uploading etc, and is seen as follows:

1. To implement the downloading of system control command in the control software, including the configuration of system parameters, beginning-ending commands for system running and a series of testing commands.

2. To implement the automatic control in QKD process through the monitoring of system error rate.

When the error rate increases to the preset threshold, the *polarization adjustment* module will be started to feedback the error rate to a low level in real time. This process is that Alice generates H signal light or P signal light, then the voltage on each polarization controller is adjusted by the *high-voltage* module until the proportion of H:V or P:N reaches to preset value which is usually set to be 150:1. After the H and P polarization adjustment are all achieved, the module keeps the new voltage on each polarization controller and changes the signal light to be random.

When SPD count rate is greatly changed, the sifting process will be paused and delay adjustment of trigger signal of the SPD will be performed. When any system error occurs, e.g. the error rate is shown as zero or the error rate can't be adjusted to the expected low level, the key extraction will be restarted.

3. To implement the exchange of the key extraction data through the USB channel between the FPGA and the single-board-computer, and the network between the communication sides.

4. To display and save the state data uploaded by the control board in the key extraction processing.

## III. QKD SYSTEM EVALUATION

In system-running test, Alice and Bob communicate through 20 km single mode fiber. The measurement shows that, the optical signal loss is 4.5 db in the fiber while 3 db loss is in Bob's optical system. Photon light is emitted at repetition-frequency of 20 MHz, and the average photon number in each pulse is $0.6*6/8+0.2*1/8=0.475$. The count rate outputted by SPDs is close to be 1% of the frequency emitted from Alice, thus the sifting key generate rate (R) is

$$R = F \cdot P \cdot C \cdot p1 \cdot p2,$$

in which F=20 MHz, P=1%, C=0.5 comes from the fact that a half of the detected photons are discarded, p1=90% comes from 10% of the signal state date are used for the statistics on the sampling error rate, and p2=18/19 comes from 1/19 of the signals will be decoy state which don't generate sifted keys. So the sifting key generate rate is about

$20MHz*1%*0.5*0.9*18/19 = 85$ kbps.

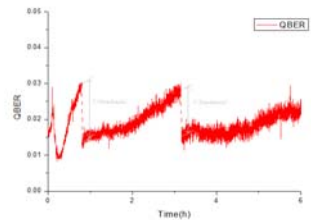

Figure 12 diagram of QBER changes with time

The quantum number error rate (QBER) changes with time in 6 hours is shown as figure 12, the threshold is set for the polarization feedback operation when system error exceeds 3%, so as to lower QBER to about 1%. The dark count of 4 SPDs is about 400/s, brings little QBER of 400/200000=0.2% and the afterpulse will bring QBER of 1/1000*0.5=0.05%, so other QBER comes from the crosstalk of polarization. Twice feedback process in the 6 hours test can be seen in the figure, and each single feedback time is nearly 1 minute.

When privacy amplification is in processing, the SFactor needs to be calculated according to the error correction situation. This function is completed in the key extraction control software and the SFactor is outputted to the *privacy amplification* module in real-time. The figure 13 shows the relationship between the SFactor and the QBER when the detected count is 1% of emitted photons. We can see that the SFactor is decreasing as the QBER increases in the same optical condition, making the final key generation rate decrease as the QBER increases. The final key generation rate change via time in 6 hours is shown as figure 14; the average rate is about 17 kbps.

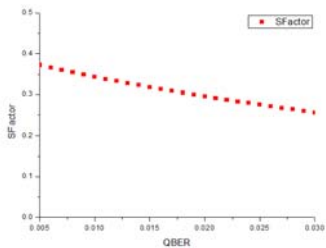

Figure 13 diagram of SFactor changes with QBER

The final key application test has also been executed by outputting the final keys to the *key management and application* module from the control board, and connecting two internet phones with the *key management and application* module. We encrypt and decrypt the voice data by XOR them with the final keys and achieve clear voice communication between Alice and Bob.

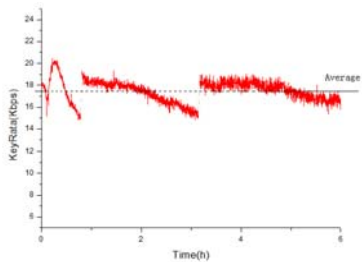

Figure 14 diagram of final key generation rate changes with time

## IV. CONCLUSION

We have successfully completed a QKD system based on polarization coding and decoy state protocol. The system completes a real-time key extraction algorithm under the control of key extraction *control* module. Meanwhile we design auxiliary modules to ensure the continuously steady and robust key generation, including TRNG, information synchronization mechanism, delay adjustment and polarization adjustment modules etc.

Finally, we statistically analyze the QBER and final key generation rate for the running system and test the system's stability with 6-hour non-stop operation. Moreover, an encrypted voice communication has been achieved.